\documentclass[universe,article,accept,pdftex,moreauthors]{Definitions/mdpiedited} 

\firstpage{1} 
\pubvolume{1}
\issuenum{1}
\articlenumber{0}
\pubyear{2025}
\copyrightyear{2025}
\datereceived{21 February 2025} 
\daterevised{24 March 2025} 
\dateaccepted{4 April 2025} 
\datepublished{ } 
\hreflink{https://doi.org/} 

\usepackage{upgreek}

\def\kms{km~s$^{-1}$}
\def\cm3{cm$^{-3}$}

\newcommand\hi{$\textrm{H}\scriptstyle\mathrm{I}$}
\newcommand\CII{$\textrm{C}\scriptstyle\mathrm{II}$}
\newcommand\cii{[\CII]}
\newcommand\nii{$\textrm{N}\scriptstyle\mathrm{II}$}
\newcommand\sii{$\textrm{S}\scriptstyle\mathrm{II}$}

\newcommand{\simlt}{\lower.5ex\hbox{$\; \buildrel < \over \sim \;$}}

\def\hl{} 
\def\highlighting{} 

\Title{Extraplanar [C II] and H$\upalpha$ in the Edge-On Galaxy NGC 5775}

\TitleCitation{Extraplanar \mbox{[C II]} and H$\upalpha$ in the Edge-On Galaxy NGC 5775}

\Author{
\hl{William T. Reach} 
 $^{1,}$\hl{*}\orcidA{}, 
Dario Fadda $^{2}$,
Richard J. Rand $^{3}$ and
Gordon J. Stacey $^{4}$
}

\AuthorNames{William T. Reach, Dario Fadda, Richard J. Rand and Gordon J. Stacey}

\AuthorCitation{Reach, W.T.; Fadda, D.; Rand, R.J.; Stacey, G.J.}

\address{%
$^{1}$ \quad Space Science Institute, 4765 Walnut Street, Suite 205, Boulder, CO 80301, USA\\
$^{2}$ \quad Space Telescope Science Institute, 3700 San Martin Dr, Baltimore, MD 21218, USA; \hl{dfadda@stsci.edu} 
\\
$^{3}$ \quad Department of Physics and Astronomy, University of New Mexico, 800 Yale Blvd NE, \linebreak  Albuquerque, NM 87131, USA; rjr@unm.edu\\ 
$^{4}$ \quad Astronomy Department, Cornell University, Ithaca, NY 14853, USA; stacey@astro.cornell.edu
}

\corres{Correspondence: wreach@spacescience.org}

\abstract{
Spiral galaxies are thin and susceptible to being disrupted vertically. 
The largest star clusters, and nuclear starbursts, generate enough energy
from winds and supernovae to send disk
material to the halo.
Observations of edge-on galaxies allow for the clearest view of vertical disruptions. We present new observations of the nearby, edge-on galaxy NGC 5775 with SOFIA \highlighting{in} 
 \cii\ 157.7 $\upmu$m
and archival images from Hubble in \highlighting{H$\upalpha$} 
 to search for  
extraplanar gas.
The extraplanar \cii\ extends 2 kpc from the midplane over much
of the star-forming disk. 
The extraplanar \cii\ at 2 kpc from the midplane approximately follows the rotation 
of the disk, 
with a lag of approximately 40 \kms; this lag is similar to what has
been previously reported in H$\upalpha$.
Significant vertical extensions (to 3 kpc) are seen 
on the northeast side of the galaxy, potentially due to super star clusters in the NGC 5775 disk combined with gravitational interaction with
the  companion galaxy NGC 5774.
The H$\upalpha$ narrow-band image reveals a narrow plume that extends 7 kpc from the nucleus and is almost exactly perpendicular to the disk. The plume
shape is similar to that seen from the comparable galaxy NGC 3628 and may arise
from the nuclear starburst. Alternatively, the H$\upalpha$ plume could be a relic of past activity.
}

\keyword{galaxies; NGC 5775; far-infrared spectroscopy}

\begin{document}

\section{Introduction}

Galaxies are composed of dark matter, stars, and~interstellar gas. In~the solar neighborhood of the Milky Way, these constituents comprise 42\%, 31\%, and~14\% of the mass surface density, respectively \citep{mckee_stars_2015}. 
The fact that these densities are in the same order of magnitude suggests a significant mutual influence among the constituents. 
Since two-thirds of the baryonic material in the solar neighborhood is in stars, the~net star formation efficiency cannot be higher than this.
In fact, the~efficiency of star formation is significantly lower (\hl{<1\%,} 
 \citep{utomo_star_2018}), requiring a mechanism that regulates
star formation and prevents all gas forming stars or being lost. Otherwise, the Milky Way would be like a globular cluster, dominated by old stars and nearly gas free \citep{renaud_star_2018}. On~scales of kiloparsecs (kpc) and larger, two disruptive mechanisms contribute: 
gravitational interactions with other galaxies, and combined kinetic energy from localized bursts of star formation. The~effects of galaxy interactions are
known from observing groups and clusters \citep{kim_galaxy_2009}. 
Galaxies not in groups or clusters, and~with no recent interaction with large 
nearby galaxies, require an internal feedback source to prevent total star~formation.

The circumgalactic medium of a galaxy depends on a balance between infall, due to the pull of dark matter and stars toward the midplane, and~outflow, due to the kinetic energy imparted by supernovae and outflows from clusters in the disk
(\hl{reviewed~by}~\citep{faucher-giguere_key_2023}). 
The cycling of material from the disk of a galaxy is a key process for understanding galaxy evolution, but~simulations yield divergent views of whether star formation feedback extends beyond halos or actually cycles through a circumgalactic medium \citep{wright_baryon_2024}. 
Bursts of localized star formation drive material into the halo through `chimneys'
\citep{heiles_radio_1996}. In~balance with infall, the~cycling is 
referred to as a `galactic fountain' \citep{bregman_galactic_1980,norman_disk-halo_1989}.
However, if~the kinetic energy imparted to the outflowing material is sufficiently large, it escapes as a `superwind'~\citep{heckman_nature_1990}.

Edge-on galaxies afford the best view of inflow and outflow from the disk.
In this paper, we present observations of the 
edge-on spiral galaxy NGC 5775, which is part of the Virgo Cluster, at~a distance of 17.4 Mpc \citep{sorceSpitzerGalaxyPhotometry2014}.
The neighboring galaxy, NGC 5774, is evidently connected to NGC 5775 by an \hi\ bridge \citep{irwin_arcs_1994}, 
indicating that the two galaxies are undergoing a minor merger. 
The  \hi\ disk has a larger-than-normal scale height due to the merger \citep{zheng_h_2022}.

Multiple lines of evidence point to a large-scale outflow from NGC 5775,
with vertical excursions from the midplane of 7 kpc in H~I \citep{lee2001} and 13 kpc in [\nii] \citep{rand2000}. In~X-rays, a~halo of hot gas is evident, potentially forming
a ``limb-
brightened outflow cone'' \citep{tullmann_multi-phase_2006}. 
Distinguishing  between individual supershells or filaments and a galaxy-wide flow is challenging, as~some extraplanar features
appear to be associated with in-disk star formation \citep{li_chandra_2008}. 
Radio continuum observations tracing synchrotron radiation, which arises from cosmic rays interacting with the large-scale magnetic field of the galaxy, 
reveal organized vertical magnetic field structures 
\citep{heald_chang-es_2022}
that are distinct from the tightly wound magnetic field in the midplane of a galaxy.
NGC 5775 is relatively bright in X-rays, consistent with other high star formation Virgo galaxies and unlike galaxies that appear more faintly in 
X-rays, possibly due to ram pressure stripping of their halos
\citep{hou_x-ray_2024}.

\begin{figure}[H]
    \includegraphics[width=.6\textwidth]{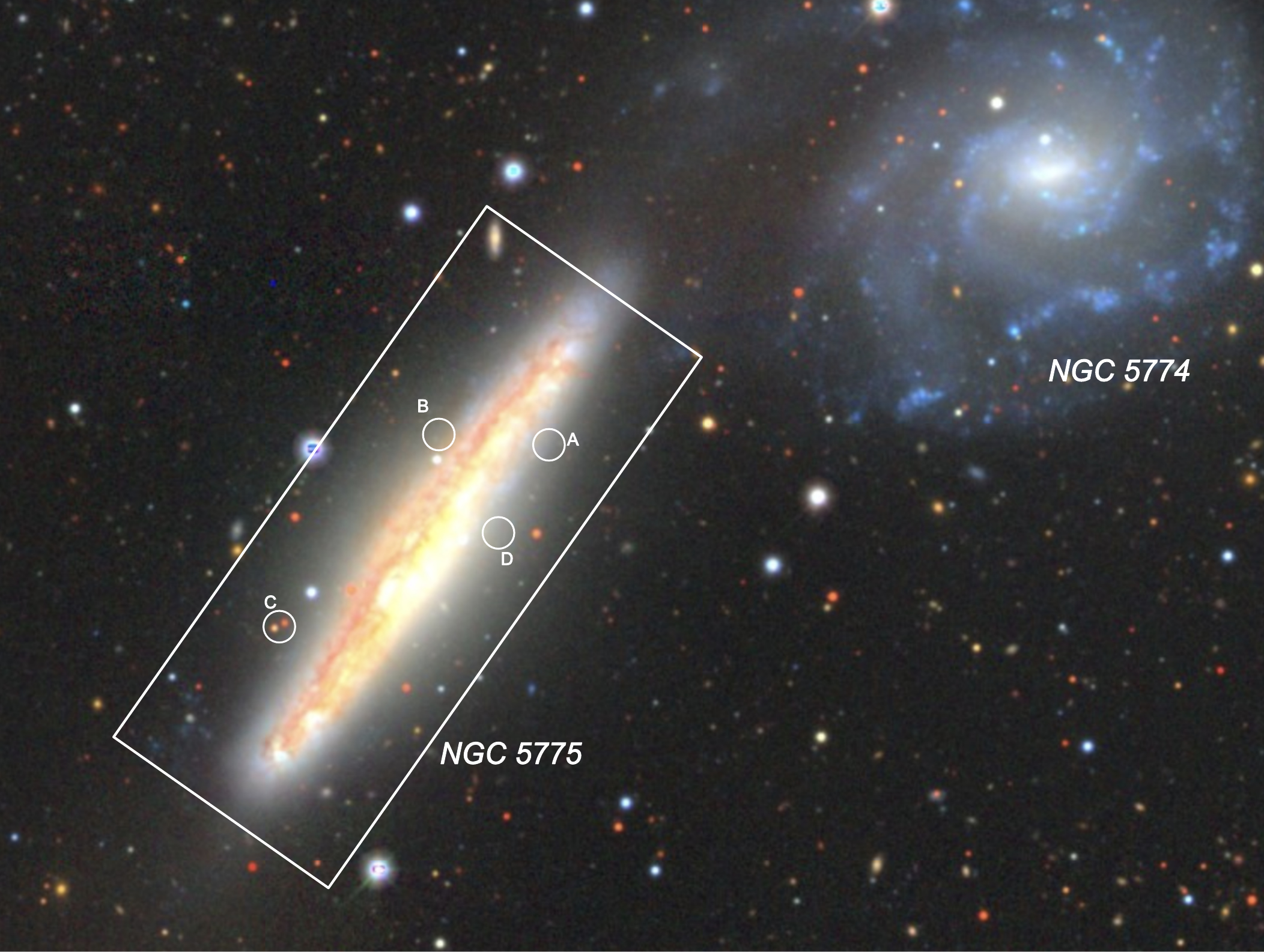}
    \caption{\hl{Sloan Digital Sky}  
Survey \citep{yorkSloanDigitalSky2000} {\it ugr} color image of edge-on galaxy NGC 5775
and face-on NGC 5774 (upper right). North is up and east is to the left. The~rectangle delineates the
coverage by SOFIA/FIFI-LS on flights 843 and 844; flight 845 avoided
the midplane itself and added integration away from the midplane. 
White circles show the locations of extraplanar regions discussed in
\hl{Section}~\ref{sec:chimneys}; 
the circle size matches the angular resolution 
of SOFIA in the \cii\ 157.7 $\upmu$m line. 
}
    \label{fig:finder}
\end{figure}

This paper completes our SOFIA observational study of edge-on galaxies,
adding new observations of NGC 5775 (Figure~\ref{fig:finder}) 
to complement those 
obtained for NGC and NGC 5907~
\citep{reach_extraplanar_2020}.
The observations target the far-infrared fine-structure line of ionized
carbon (\cii) at 157.751~$\upmu$m. The~upper energy level of this transition is
only 91 K above the ground state, making it relatively easily excited, 
except in the coldest interstellar conditions. Carbon is also easily
ionized (singly) under most interstellar conditions, except when~deep in 
cold molecular clouds (where the upper level is not excited anyway) and very close  to energy sources.
The combination of its low-excitation energy, typical ionization state, and~far-infrared emission's relative 
immunity to the heavy extinction near the galactic midplane makes 
the \cii\ ground-state fine-structure line an ideal tracer of the
diffuse interstellar~medium.

\unskip

\section{Observations}

The Stratospheric Observatory for Infrared Astronomy (\hl{SOFIA;} \citep{young_early_2012}) observed NGC 5775 as part of project 06\_0010, 
on three flights operating from Santiago, Chile.
Table~\ref{obstab} summarizes the observing conditions. 
The atmospheric transmission, averaged within $\pm 250$ km~s$^{-1}$ of the 
\cii\ 157.741 $\upmu$m line redshifted to 1676 km~s$^{-1}$,
exceeded 89\%, which is only possible from an airborne or space telescope due to
the high opacity of the Earth's atmosphere in the far-infrared.
The science instrument utilized was the Far Infrared Field-Imaging Line Spectrometer 
(\hl{FIFI-LS;} \citep{fischer_fifi-ls_2018}), which generates two images of $5\times 5$ spaxels: one red and one blue. 
The red-side grating was scanned over the wavelength of the \cii\ $^2$P$_{3/2}\rightarrow ^2$P$_{1/2}$ transition redshifted to 1676 km~s$^{-1}$.
The blue-side grating was scanned over the wavelength of the 
[O~III] $^3$P$_1 \rightarrow ^3$P$_0$ transition at 88.4 $\upmu$m;
those data are not discussed further in this paper.
The observations had an angular resolution of  $16''$, corresponding to 1.3 kpc at the distance of NGC 5775. 
The spectral resolution of the observations was 250 km~s$^{-1}$ \citep{colditz_spectral_2018}.
The observations utilized a matched chop-nod choreography, with~the secondary mirror chopping by a total separation listed in Table~\ref{obstab} as `Chop'  at a frequency of 1 Hz, with~a position angle $305^\circ$ E of N.
The exposure times in Table~\ref{obstab} comprise 37\% of the duration of the respective legs of the observing flights, with~the same amount spent on chopping using the secondary mirror for sky subtraction; the
remainder of the time was spent on telescope motions and nod~spectra. 

\begin{table}[H]
\caption{SOFIA/FIFI-LS observation~log
	.\label{obstab}}

    \begin{adjustwidth}{-\extralength}{0cm}
\begin{tabularx}{\fulllength}{cccccccccccc} 
		\toprule
\textbf{Date} & \textbf{Flight} & \textbf{Altitude (feet)} & \textbf{Elevation ($^\circ$)} & \textbf{Duration (min)} &\textbf{Chop ($''$)} & \textbf{Transmission} & \textbf{Exposure (s)}\\
		\midrule
2022 Mar 23 & 843 & 40,500 & 36 & 55  & 120 & 89 & 1443\\ 
2022 Mar 24 & 844 & 42,400 & 45 & 107 & 120 & 93 & 2488\\ 
2022 Mar 27 & 845 & 42,900 & 49 & 30  & 160 & 94 & 768 \\
		\bottomrule
\end{tabularx}
    \end{adjustwidth}

\end{table}
\vspace{-12pt}

The FIFI-LS data were processed using the SOFIA data processing pipeline
\linebreak {\tt sofia-redux}~\citep{clarke_redux_2015} version 2.5.2. Data from all flights were 
processed together, with~the following adjustments to the default parameters. 
In the Spatial Calibrate step, the~option to rotate the images by the detector angle was disabled. 
Since our observing strategy aligned the galaxy’s midplane with the detector, the~resulting images are naturally aligned with the galaxy.
In the Resample step, the~`spatial edge threshold' was decreased to 0.01 to~avoid generating artifacts from outside the boundary of the actual observed image. 
The exposure time per sample in the final cube (i.e., at~a given position and wavelength) was 72 \hl{s} 
 within 1~kpc of the midplane, 60 \hl{s} at distances of 1.5 kpc from the midplane, and~44 \hl{s} at distances of 2--3.5 kpc from the~midplane.

\begin{figure}[H]
\includegraphics[width=0.98\textwidth]{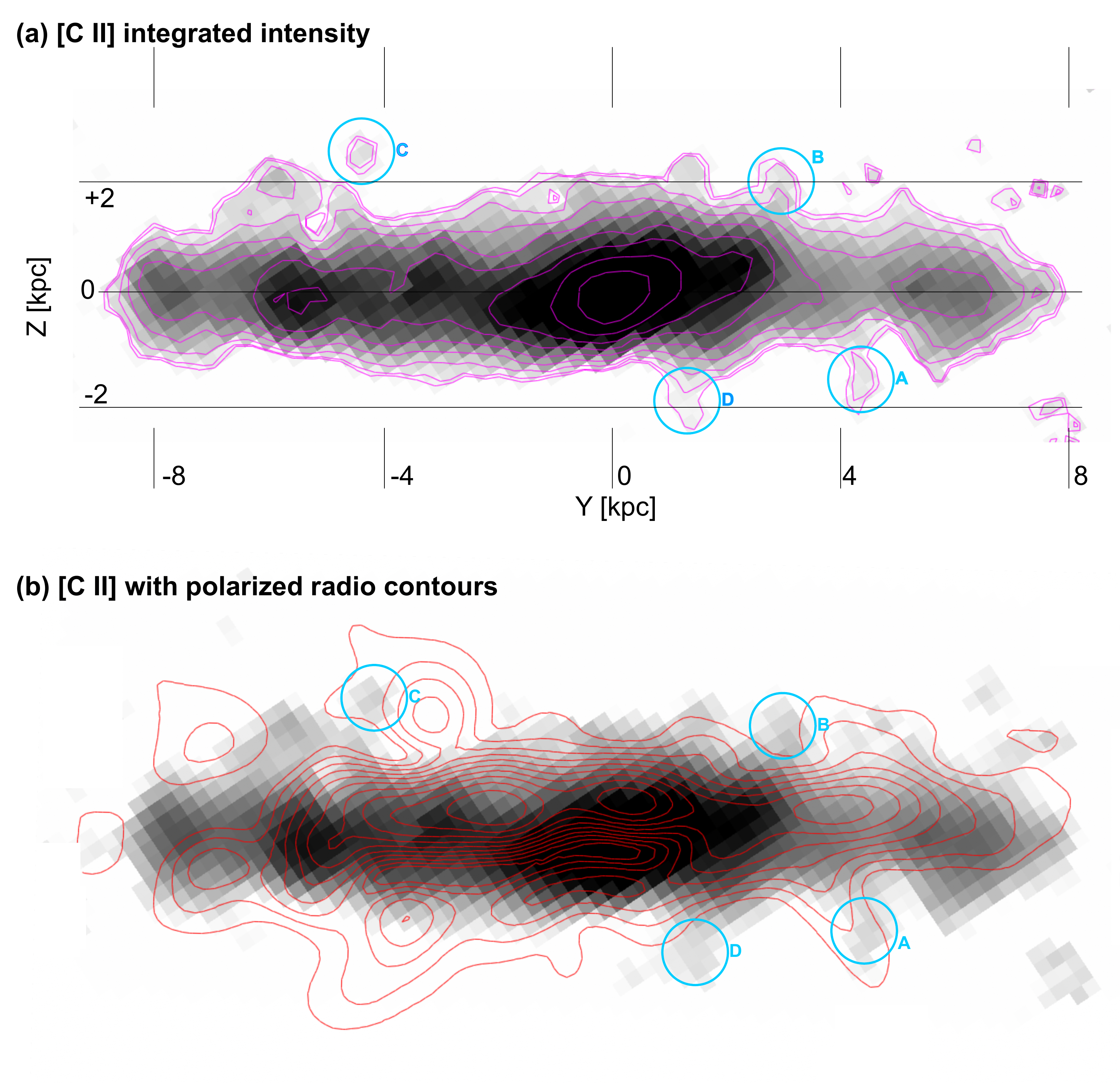}
\caption{
(\hl{\textbf{a}) SOFIA/FIFI-LS} 
 integrated \cii\ line flux image of NGC 5775, in~greyscale with contours overlaid.
The contours are at 3.0, 3.3, 4.1, 5.5, 7.4, 10, 13, 16, and~21, in~units of $10^{-7}$ W~m$^{-2}$~sr$^{-1}$. 
The $Y$ and $Z$ spatial axes are defined with vertical and horizontal lines, respectively.
(\textbf{b}) The \cii\ image in greyscale with polarized 5 GHz radio continuum \citep{irwinCHANGESXXHighresolution2019} overlaid.
Contour levels are linearly spaced from 0.014 to 0.077 K (polarized) brightness temperature.
Regions of interest A, B, C, and~D are discussed in \hl{Section}~\ref{sec:chimneys}.
    \label{fig:map}
    }
\end{figure}

\begin{figure}[H]
\includegraphics[width=0.5\textwidth]{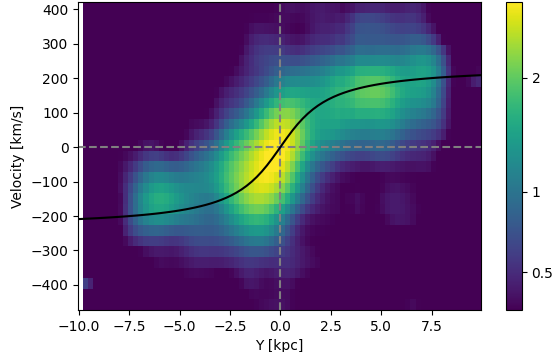}
\includegraphics[width=0.47\linewidth]{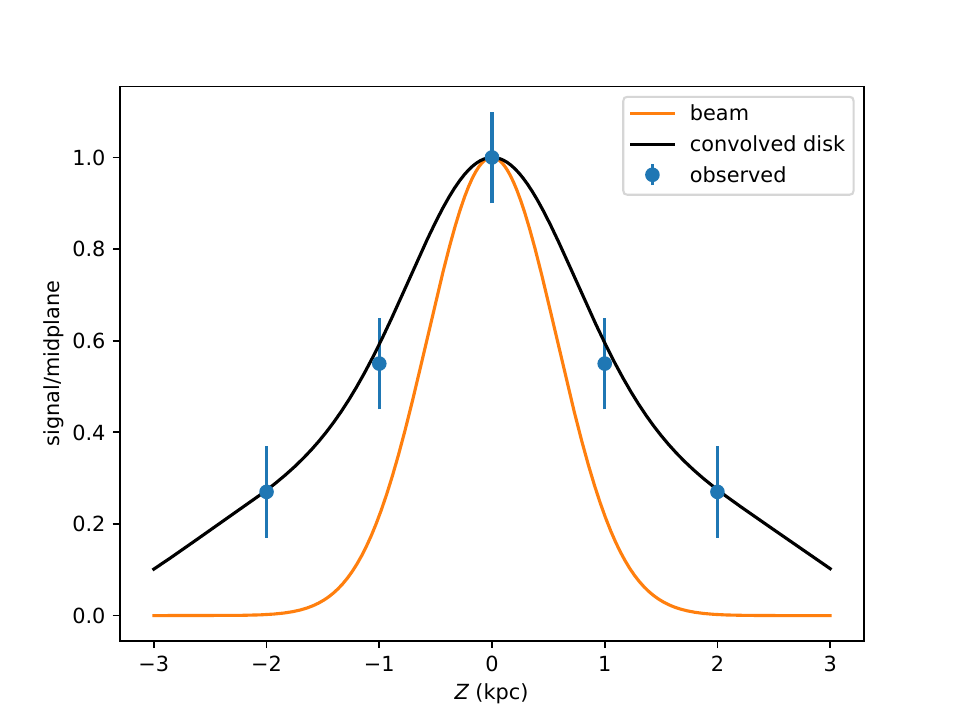}
\caption{
{(\textbf{left})} \cii\ position-velocity diagram along the major axis of the galaxy. Each pixel shows the flux per beam, with~the color bar indicating flux in Jy. A~simple rotation curve is shown in black.
{(\textbf{right})}
Vertical profile of \cii\ emission from NGC 5775, normalized to the midplane.
    The observed (circles with error bars) profile is much wider than the beam (orange curve).
    The two-scale-height disk model discussed in the text, convolved with
    the beam, is shown in black.
}    \label{fig:pv}
\end{figure}

A baseline was subtracted from each pixel of the FIFI-LS data cube by fitting a straight line to the brightness measured between $-$860 and +860 km~s$^{-1}$ (relative to the galaxy's central velocity), while excluding velocities between $-$400 and 400 km~s$^{-1}$. Figure~\ref{fig:map} shows the integrated \cii\ 
surface brightness image over the entire galaxy in the $Y,Z$ coordinate system we will use for the along-plane and vertical directions, respectively. The~left panel of Figure~\ref{fig:pv} shows a slice through the \cii\ cube along
the major axis of the galaxy. In~this and all subsequent velocity plots, 
velocities are shifted such that $v=0$ corresponds to a heliocentric systemic velocity of 1590 \kms.
An approximation of the \cii\ rotation curve in this reference frame is
\begin{equation}
    v = 150 \tan^{-1}\left(\frac{Y}{1.8\,{\rm kpc}}\right) \,{\rm km~s}^{-1}. \label{eq:vrot}
\end{equation}
This approximation is illustrated in Figure~\ref{fig:pv}.

\vspace{-9pt}
\unskip

\section{Distribution of Extraplanar [C II]  \label{sec:distribution}}
\unskip

\subsection{Vertical~Profile}

\begin{figure}[H]
\includegraphics[width=0.33\textwidth]{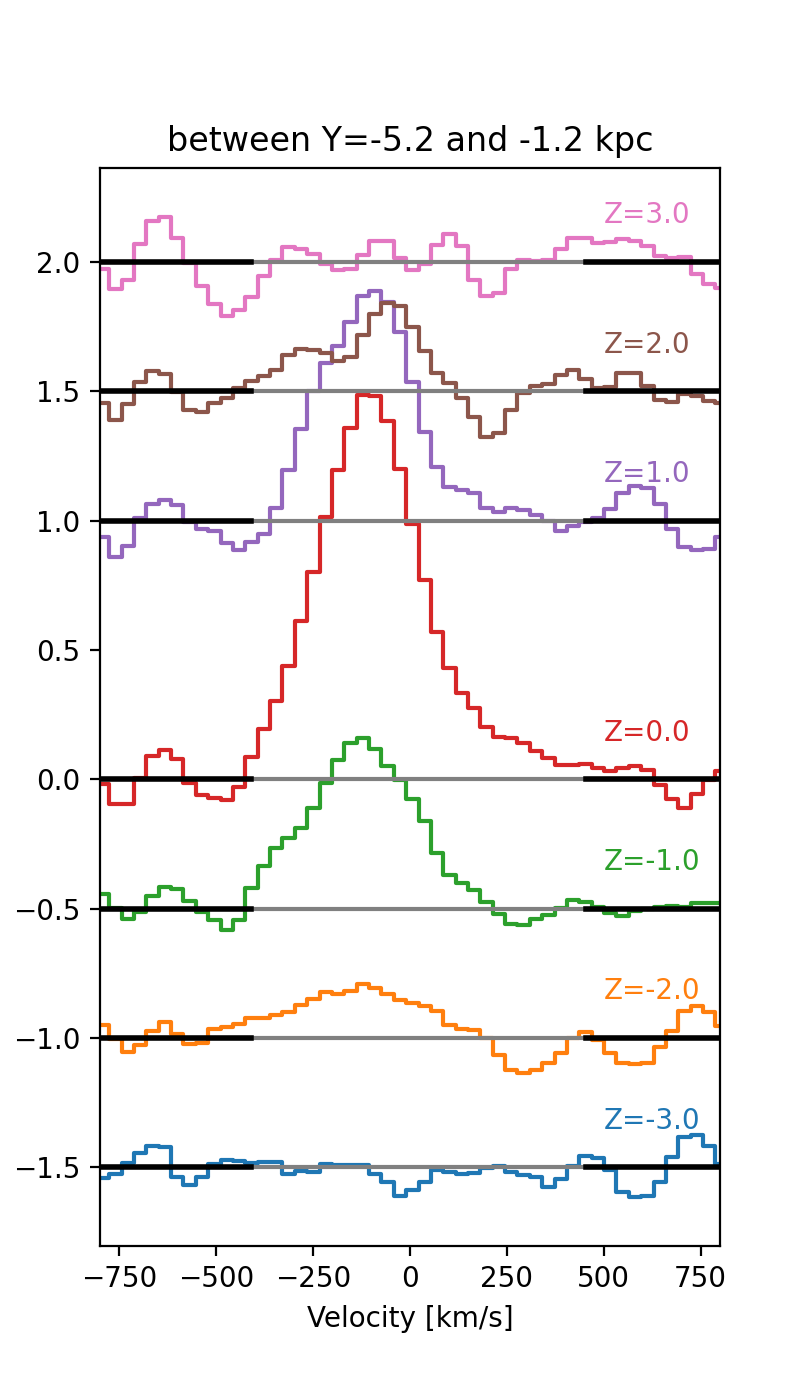}
\includegraphics[width=0.33\textwidth]{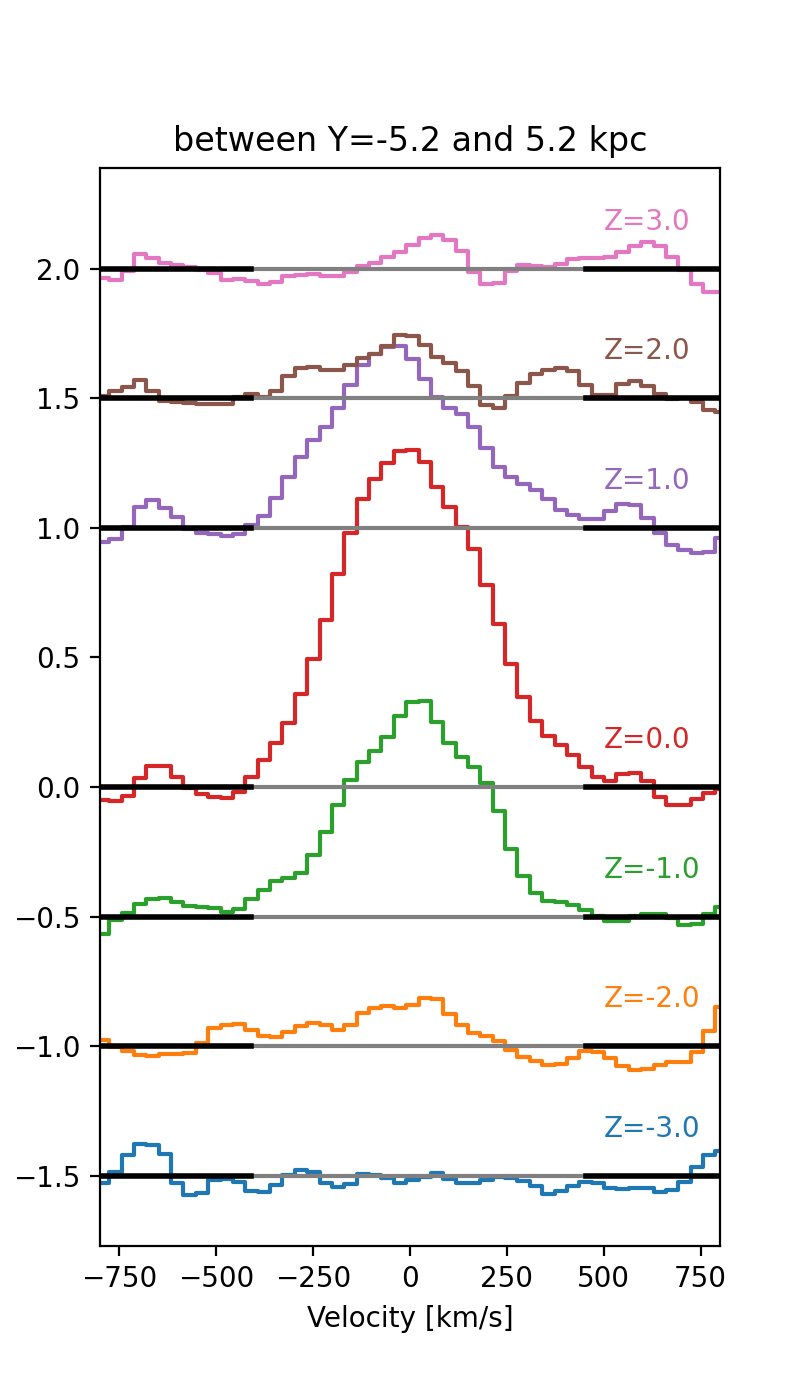}
\includegraphics[width=0.33\textwidth]{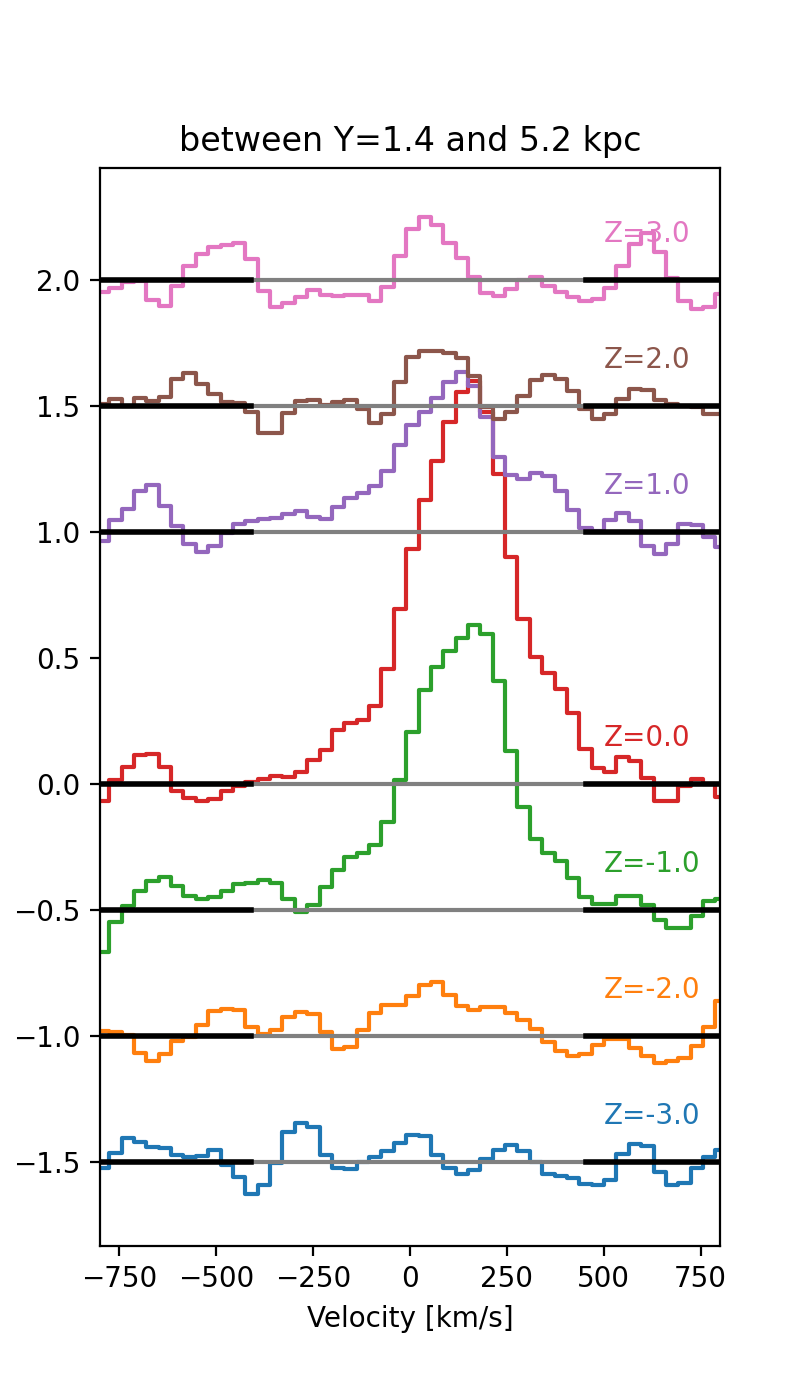}
\caption{\cii\ \hl{spectra of} 
 NGC 5775 at different heights above and below the midplane. Each spectrum is labeled by its $Z$ height and is offset vertically for clarity, with~the zero-intensity line shown in grey and baseline region in black. The~middle panel averages over most of the disk ($-$5.2 to 5.2 kpc). The~left and right panels show the spectra averaged over the approaching and receding parts of the disk, respectively, at~distances greater than 1.2 kpc from the nucleus.  
}
    \label{fig:spec}
\end{figure}

To derive the vertical extent of the \cii\ emission, we extracted spectra
at various altitudes ($Z$) above the midplane, averaging over elongated rectangular regions within one beam of each $Z$. 
Since the galaxy's rotation speed
exceeds the instrument's velocity resolution, averaging the spectra over the northern and southern halves of the galaxy
would artificially dilute the signal.
To mitigate this effect, we created three sets of vertical spectra, each corresponding to a different range of projected distances ($Y$) from the nucleus, measured parallel to the midplane.
Figure~\ref{fig:spec} shows the spectra for the approaching and receding
halves of the galaxy, as~well as the average over the~disk.

The \cii\ emission decreases steeply with $Z$, but~its vertical distribution remains well-resolved. 
The SOFIA full width at half maximum (FWHM) beam is 1.35 kpc at the distance of NGC 5775, corresponding to a spatial extent of 
$\pm 0.67$ kpc. Emission is clearly detected at both $Z=+2$ and $Z=-2$ kpc from the midplane. 
The telescope is diffraction-limited at the observing wavelength.
For a Gaussian approximation of the beam, and~a disk 
significantly thinner than the beam, the~signal of the midplane 
that an observer would detect at separation $Z$ = 0, 1, and~2 kpc would be 100\%, 22\%, and <0.3\%
of its peak signal. This applies to a disk with a scale height much smaller than the beam, 
$H<100$ pc.
For a thicker disk with an exponential scale height, $H=0.5$ kpc, the~signal that would be observed 
at 1 and 2 kpc above the midplane would be 45\% and 0.07\% of the midplane brightness, respectively. This scale
height approximately matches the observed brightness at 1 kpc but is
4 times too low to explain the brightness at 2~kpc.

Thus, the \cii\ emission does not arise solely from the thin disk of star formation
in the midplane. This same result was obtained for NGC 891 and NGC 5907,
where a two-scale-height fit was found to have a thinner disk of 0.4 kpc
and a thick disk scale height of 2.8 kpc \citep{reach_extraplanar_2020}.
Using those two scale heights, the~variation of \cii\ with $Z$ in NGC 5775 can be reasonably well fit,
suggesting the vertical distribution is similar.
By fixing the thinner disk scale height to 0.4 kpc and fitting the vertical profile
as shown in Figure~\ref{fig:pv} (right),
we derive a thick disk scale height of $H_2=2.3\pm0.7$ kpc for NGC~5775.

\def\extra{
\begin{figure}[H]
    \centering
    \includegraphics[width=0.99\linewidth]{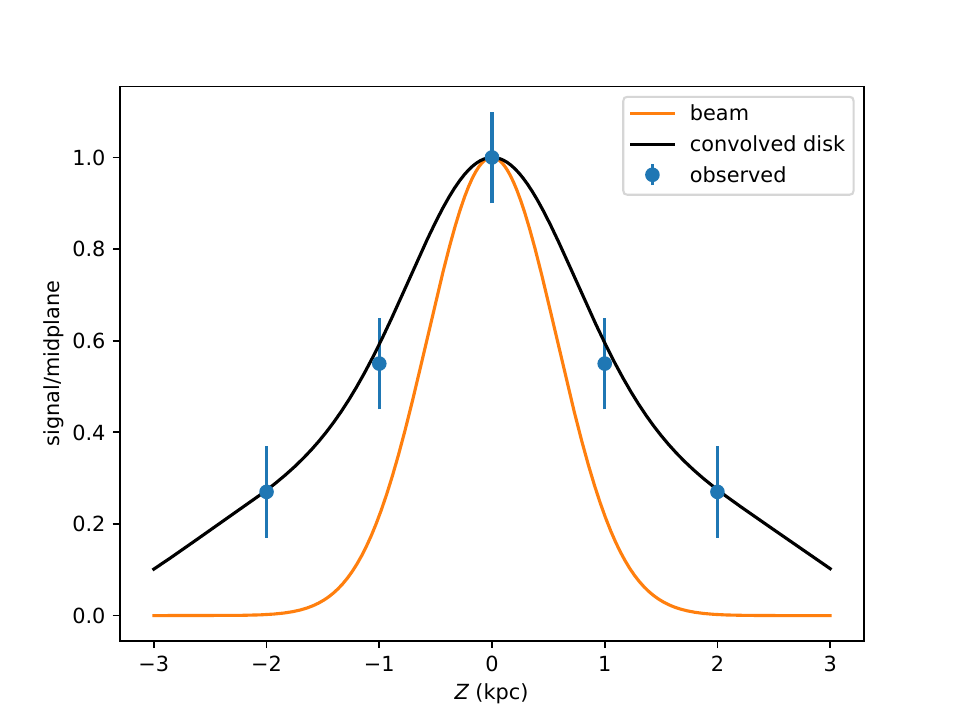}
    \caption{Vertical profile of \cii\ emission from NGC 5775, normalized to the midplane.
    The observed (circles with error bars) profile is much wider than the beam (orange curve).
    The two-scale-height disk model discussed in the text, convolved with
    the beam, is shown in black.
    \label{fig:vertical}}
\end{figure}
}

The diffuse ionized gas is significantly extended on the northeast side of the galaxy, beyond~what can be explained by 
thermal, magnetic, turbulent, or~cosmic ray pressure~
\citep{boettcher_dynamical_2019}.
A possible explanation is that the halo gas is aligned with the direction toward the nearby major galaxy NGC 5774, which lies west of NGC 5775 (see \hl{Figure}~\ref{fig:finder}) 
and exerts greater gravitational influence on the negative $Z$ side of the galaxy. 
\citet{irwin_arcs_1994} identified an \hi\ bridge extending from the northern side of NGC 5775 to NGC 5774,
connecting the galaxies and originating (based on the observed radial velocity) from the latter galaxy. 
The \hi\ bridge and associated radio continuum emission form an envelope encompassing both galaxies~\citep{lee2001}.

\subsection{Rotation of Extraplanar~Gas}
The extraplanar \cii\ approximately follows the rotation of the disk, but~with a systematic slowing away from the midplane.
On the northeastern side of the galaxy, the~line center shifts to a lower line-of-sight velocity by 
$40\pm 8$ \kms\ at $|Z|=2$ kpc, corresponding to a gradient of $20\pm 4$ \kms\ per kpc.
The apparent lag seen in \cii\ is similar in sign and magnitude to 
the velocity lag observed in H$\upalpha$ on the northeastern side of the galaxy.  
The lag represents a small fraction of the 150 \kms\ amplitude of the rotation curve (Equation~(\ref{eq:vrot})) and is below the 
250 \kms\ velocity resolution,
so it is only measured statistically with the SOFIA data.
Part of this apparent lag could be caused by line-of-sight geometrical effects, as~the galaxy is not perfectly edge-on,
and the \cii-emitting material may not be uniformly distributed along the line of sight.
For comparison, the~H$\upalpha$ emission has a clear velocity gradient along the minor axis of 25 \kms\ per kpc on the northeast side and 
6 \kms\ per kpc on the southwest side of the galaxy
\citep{boettcher_dynamical_2019}, while an earlier study found an
average gradient of 8 \kms\ per kpc \citep{heald_imaging_2006}.
For stars, vertical gradients of azimuthal velocity may be due to the
influence of a spheroidal mass distribution 
\citep{jalocha_vertical_2011}. 
For the gas, models of ejected clouds following
ballistic trajectories show some resemblance to the observed
trends but cannot fully explain the kinematics, which may be 
significantly affected by the interaction with NGC 5774
\citep{collins_kinematics_2002}.

\subsection{Potential Chimneys and Extraplanar~Clouds  \label{sec:chimneys}}

\begin{figure}[H]
\includegraphics[width=0.5\textwidth]{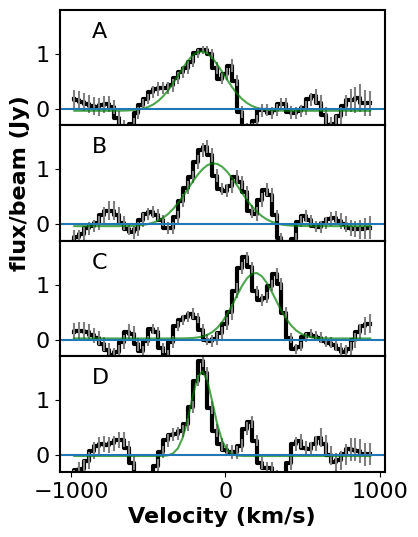}
\caption{
\cii\ spectra 2 kpc above potential midplane chimneys of NGC 5775, 
as labeled in Figure~\ref{fig:finder}.
For each of these positions, the~average of spectra at the same vertical distance from the midplane, with a 1.5-beam offset on either side
, was subtracted.
Gaussian fits to the velocity distributions are shown as green curves, with~the baselines as straight blue lines.
}
    \label{fig:specExtra}
\end{figure}

Local deviations exist in the overall vertical distribution of extraplanar gas. 
Notably, \cii\ emission is observed even at 3 kpc above the midplane on the northeast side of the galaxy (positive $Z$ and positive $Y$),
and there are some spots of enhanced brightness relative to their surroundings.
Figure~\ref{fig:specExtra} shows  \cii\ deviation spectra for four individual regions (the~locations of which are labeled in
Figures~\ref{fig:finder} and~\ref{fig:map}). The~deviation spectra were calculated for each position by subtracting the average of reference 
spectra at the same $Z$ but offset by 1.5~beam widths in $\pm Y$.

Position A is on a prominent radio polarization spur, and~it shows a broad velocity distribution, with~FWHM 350 \kms and center $-$150 \kms. 
The contours of \cii\ and radio polarization closely agree. This is the best example of a potential \cii\ in NGC 5775.
Position B is on a \cii\ spur with no corresponding radio polarization feature. Its spectrum has a broad distribution like position A,
but with the center shifted to $-$80 \kms, primarily due to galactic rotation (as this position is closer to the nucleus). 
The small-scale structure in the spectrum of position B is finer than the instrumental resolution of 250 \kms, but~the centroid shift
is well-measured and indicates deviations from symmetric rotation at $+Z$ and $-Z$. 

Position C is extraplanar peak of \cii\ brightness detached from the disk of NGC 5775. 
It has a somewhat narrower \cii\ velocity distribution, with~FWHM 300 \kms. The~ centroid at +190 \kms clearly shifted from 
the other spectra due to galactic rotation (as this position is on the opposite side of the galaxy from the others).  

\textls[-15]{Position D is another \cii\ spur, without a~clear radio counterpart. The~\cii\ velocity distribution here is
narrow, with~FWHM 293 \kms\ primarily due to the instrumental~resolution.}

\section{Large-Scale Feedback from the Nuclear~Region}

The far-infrared surface brightness of the nuclear region is significantly elevated, indicating a high star formation rate from the inner portions of NGC 5775's disk. 
The concentrated star formation can lead to a nuclear starburst and superwind, as seen in M~82 \citep{heckman_nature_1990}. Such winds are broad, extending symmetrically on both sides of the galaxy with a `U'-shaped opening.  
The central black hole can also release vast amounts of energy into its surroundings when material is accreted. 
Accretion disks generally drive jets 
 parallel to their spin axis, as~they must shed angular~momentum. 

While there is no evidence of a nuclear superwind from the far-infrared continuum or \cii\ observations, the~ionized extraplanar gas far from the disk is traced using optical H$\upalpha$ emission. 
NGC 5775 was among the galaxies in studies that showed the extent and frequency of galactic winds potentially driven by supernovae \citep{lehnertIonizedGasHalos1995,lehnertIonizedGasHalos1996}.
H$\upalpha$ images \citep{collins_diffuse_2000} show vertical filaments from the disk and compared to models of ballistic ejection \citep{collins_kinematics_2002}. 
NGC 5775 was included in a systematic study of extraplanar H$\upalpha$ emission \citep{rossaHaSurveyAiming2003,rossaHaSurveyAiming2003a}, where prominent extraplanar features were 
detected in both H$\upalpha$ and~dust.

In this section, we  search for nuclear feedback to~complement the SOFIA \cii\ observations presented earlier in this paper, while not repeating the existing H$\upalpha$ studies on extraplanar gas. 
We use archival Hubble Space Telescope images of NGC 5775, which were obtained as part of observation 
 project 10416 (PI J. Rossa). 
A comparison of these Hubble data to Chandra has been made \citep{li_chandra_2008}.
The Hubble Legacy Archive continuum and H$\upalpha$ narrow-band filter images were 
reprojected onto a 
common grid. The~ratio of H$\upalpha$ to continuum filter brightness for starlight was used to scale and subtract the continuum from the H$\upalpha$~image.

\begin{figure}[H]
\includegraphics[width=0.98\textwidth]{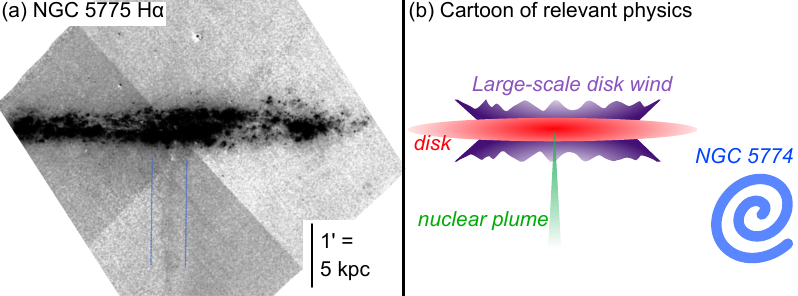}
\caption{
{(\textbf{a})} H$\upalpha$ image of NGC 5775, showing a faint nuclear jet to the south (within the two vertical blue lines).
{(\textbf{b})} Simplified cartoon showing the various geometrical structures mentioned
in the Conclusions, approximately to geometric scale.
    \label{fig:hstjet}
    }
\end{figure}

Figure~\ref{fig:hstjet}a shows that a large-scale (>7 kpc) nuclear plume extends southward from the galaxy, directly below the nucleus.
A  counterplume may also be present in the images, though~it is fainter.
The plume closely resembles,~both in size and morphology, the~one found in NGC 3628, which was
interpreted as being part of a nuclear starburst wind outflow.
Deep optical spectra show that the
nuclear plume in NGC 3628 has high ratios of
[\nii]/H$\upalpha$ and [\sii]/H$\upalpha$, suggesting shock
origin \citep{fabbiano_nuclear_1990}. Alternatively, the~observed line ratio could
be due to a dilute radiation field at high $Z$.

The plume is clearly distinct from the larger-scale disk wind in~that it is much more
collimated. From~its morphology alone, the~plume resembles a jet, similar to those produced
by supermassive black holes in active galactic nuclei. 
Although there is no known evidence
for an active nucleus in NGC 5775, this does not rule out past activity. The~H$\upalpha$ plume could be a relic of a prior active~jet. 

The H$\upalpha$ surface brightness of the plume in the HST image is approximately 
\mbox{$5\times 10^{-7}$ erg~cm$^{-2}$~s$^{-1}$~sr$^{-1}$}, measured by making a slice 
through the plume parallel to the midplane and 5 kpc below it. 
This corresponds to an emission measure \mbox{EM $=6 T_4^{0.92}$ cm$^{-6}$~pc},
where $T_4$ is the gas temperature in units of $10^4$ K \citep{reynolds1992}.
There is no evidence of the plume in high-resolution
radio images taken with the Very Large Array 
(\hl{VLA;} \citep{irwinCHANGESXXHighresolution2019}). 
The predicted radio brightness temperature of the plume at 5 GHz
is $T_B=0.3 T_4^{-.35}$ mK.
The VLA resolution at this frequency is $12''$ and dilutes the plume, which has a diameter $\sim 7''$, further 
reducing the observable brightness temperature.
An approximate upper limit from the VLA images is 
$T_B<3$ mK, which is not enough to detect this faint~structure.

The emission measure is defined as the line-of-sight integral EM $\equiv\int n_{\rm H}^2 f^{\frac{1}{3}}{\rm d}L$, assuming the hydrogen is fully ionized,
and has a local density $n_{\rm H}$ and~a 
volume filling factor $f$.
The  gas density is then 
$n_{\rm H}\sim 0.1 T_4^{0.46} f^{-1/6}$ cm$^{-3}$,
over the line-of-sight path length $L\sim 700$ pc.
Approximating the plume as a cylinder, its total mass \hl{is} 
$6\times 10^7 \mu n_{\rm H} f^{\frac{5}{6}}$ $M_\odot$, where $\mu\sim 1.4$ is the mean weight per hydrogen, including helium and heavier elements.
The total rate of ionizing photons required to keep the plume ionized,
using the case B recombination coefficient \citep{draine2011}, 
is $2\times 10^{50} f^{\frac{5}{6}}$ s$^{-1}$.
This much ionizing radiation could be provided by as few as four extraplanar O5-type stars or 100 of the more common 
 O9-type stars. 
However, the~present ionized state of the plume does not require extraplanar
young stars. Instead, the~material could have been injected into the halo at a
high temperature, already ionized. The~recombination time, using the density estimated above, is $10^4 f^{\frac{1}{2}}T_{\rm init}$ \hl{yr}, 
 where $T_{\rm init}$ is the peak temperature.
If the material had been ejected in a high-energy jet, at a~temperature of $10^8$ K, 
it would remain ionized for $10^8$ \hl{yr}. 
Thus, the~plume may be a relic of a 
past active phase of the supermassive black hole at the center of the galaxy,~appearing in a jet-like form 
despite the current absence of AGN activity in NGC~5775.

\section{Conclusions}

\def\extra{
\begin{figure}[H]
    \centering
    \includegraphics[width=.75\textwidth]{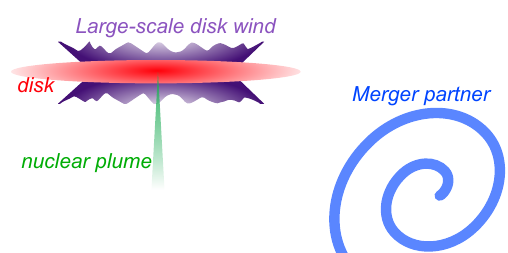}
    \caption{
Simplified carton showing the various geometrical structures mentioned
in the Conclusions, approximately to geometric scale.
}
    \label{fig:cartoon}
\end{figure}
}

The observations of NGC 5775 presented in this paper suggest
two distinct modes of mass injection into the halo, which are schematically shown in Figure~\ref{fig:hstjet}b:

(1) A galaxy-wide superwind is produced by a widespread burst of star formation, 
including the nucleus as well as super star clusters throughout the disk. 
New far-infrared observations with SOFIA detect far-infrared  \cii\ 
emission from NGC 5775 
rising up to at least 2 kpc from the midplane, similar to what was found in NGC 891 and NGC 5907 \citep{reach_extraplanar_2020}. 
The extraplanar gas distribution is more extended closest to the nearby major galaxy NGC 5774, which exerts significant
gravitational~influence.

The surface brightness of the extraplanar \cii\ at 2 kpc from the midplane of
NGC 5775 (24 nW~m$^{-2}$~sr$^{-1}$) is comparable to the northern starburst region of NGC 891 (20~nW~m$^{-2}$~sr$^{-1}$).
If the \cii\ arises from gas where the H is ionized, the~implied emission measure is 1000 cm$^{-6}$ pc, and~the gas density is of order 1 cm$^{-3}$.
Such a high gas density at 2 kpc above the midplane exceeds values predicted by state-of-the-art simulations of galactic outflows \citep{schneiderIntroducingCGOLSCholla2018}. 
If the \cii\ instead arises
from gas where the H is neutral, an~even higher volume density of order 40 cm$^{-3}$ is required, due to the lower efficiency of excitation via neutral collisions. 
Transporting such dense gas to high
altitudes above the midplane, or~forming it in~situ, presents a challenge and may have important implications for the
study of circumgalactic multi-phase~gas.

The \cii\ velocity in NGC 5775 is similar to that of other tracers of
the interstellar medium in the disk, but~it lags 
galactic rotation at high altitudes.
High-sensitivity polarized radio continuum images may trace individual conduits between the disk and halo from individual super star clusters \citep{irwin2024}.
Soft (0.2--0.5 keV) X-ray emission reveals `spurs' that extend from the midplane to 2 kpc,
with possible H$\upalpha$ and \cii\ counterparts, originating
approximately 6 kpc along the midplane from the nucleus
\citep{tullmann_multi-phase_2006}. These spurs may delineate
the walls of a galaxy-scale supershell, driven by the combined
star formation activity of the entire inner~disk. 

(2) A more collimated
plume of material, originating in the nucleus,
is detected in faint H$\upalpha$ emission. 
The plume may be a remnant of a nuclear 
jet from a supermassive black hole that was active in the last $10^8$ \hl{yr} but was not active
at the time of the HST~observation.

\vspace{6pt} 


\authorcontributions{
Conceptualization, W.T.R., D.F., R.J.R., and G.J.S.; 
formal analysis, D.F.; methodology, W.T.R.; software, D.F.; 
writing---original draft, W.T.R.; writing---review and editing, R.J.R. All authors have read and agreed to the published version of the manuscript.}

\funding{Financial support for this work was provided by the U.S. National Aeronautics and Space Administration through award \#06-0010 issued by the Universities Space Research~Association.}

\dataavailability{The SOFIA data used in this project are available from the Infrared Science \hl{Archive at} 
URL (accessed on 7 Apr 2025) \url{https://irsa.ipac.caltech.edu/applications/sofia/}. 
The~Hubble Space Telescope data used in this project are available from the Mikulski Archive for Space \hl{Telescopes at} URL (accessed on 7 Apr 2025) \url{https://mast.stsci.edu/portal/Mashup/Clients/Mast/Portal.html}.}

\acknowledgments{This study is based 
	 in part on observations made with the NASA/DLR Stratospheric Observatory for Infrared Astronomy (SOFIA). SOFIA was jointly operated by the Universities Space Research Association, Inc. (USRA), under~NASA contract NNA17BF53C, and~the Deutsches SOFIA Institut (DSI) under DLR contract 50 OK 0901 to the University of Stuttgart. This research has made use of ``Aladin sky atlas'' developed at CDS, Strasbourg Observatory, France \citep{bonnarelALADINInteractiveSky2000}.}

\conflictsofinterest{The authors declare no conflicts of~interest.} 

\begin{adjustwidth}{-\extralength}{0cm}

\reftitle{References}



%


\PublishersNote{}
\end{adjustwidth}
\end{document}